\documentclass[9pt,twocolumn,twoside]{osajnl}
\journal{optica}
\setboolean{shortarticle}{true}

\newcommand{\vE}{{\bf E}}
\newcommand{\ld}{\lambda}
\newcommand{\gm}{\gamma}
\newcommand{\Gm}{\Gamma}
\newcommand{\Dt}{\Delta}
\newcommand{\nn}{\nonumber}
\title{Extraordinary fluorescence enhancement in metal-dielectric core-shell nanoparticles}

\author[1,*]{Ilia L. Rasskazov}
\author[2]{Alexander Moroz}
\author[1]{P. Scott Carney}

\affil[1]{The Institute of Optics, University of Rochester, Rochester, NY 14627, USA}
\affil[2]{Wave-scattering.com (e-mail: wavescattering@yahoo.com)}
\affil[*]{Corresponding author: irasskaz@ur.rochester.edu}

\begin{abstract}
Contrary to a paradigm of metal-enhanced fluorescence, unusually thick dielectric coatings can be very favorable to achieve extreme values of averaged fluorescence enhancement factor $\bar F\gtrsim 3000$ for emitters located on the surface, or in the interior, of the shell of Au@dielectric spherical core-shell particles under realistic conditions, even for the emitters with 100\% intrinsic quantum yield.
Thick dielectric coatings facilitate high-quality transverse electric (TE) multipole ($\ell=7$) resonances which are shown as the major cause for the reported extraordinary values of $\bar F$.
\end{abstract}

\setboolean{displaycopyright}{true}

\begin{document}

\maketitle

%%%%%%%%%%%%%%%%%%%%%%%%%%%%%%%%%%%%%%%%%%%%%%%%%
Plasmonic nanostructures potential in fluorescence enhanced emission has been examined over long time~\cite{Ford1984,Geddes2002,Moroz2005,Moroz2005a,Tovmachenko2006,Anger2006,Bharadwaj2007,Arruda2017a}. 
Collective electron oscillations on the surface of plasmonic nanostructure can generate a strong local electric field, $\vE$, enhancement to boost the excitation rate, $\gamma_{\rm exc}$ ($\propto |\vE|^2$), of a fluorophore.
At the same time, reducing the distance of a fluorophore from a metal nanostructure opens dissipative channels which diminish the quantum yield, $q$, and negatively affect fluorescence. An optimal {\em averaged} fluorescence enhancement factor,
\begin{eqnarray}
\lefteqn{
 \bar F = \dfrac{\gm_{\rm exc}}{\gm_{\rm exc;0}} \dfrac{q}{q_0} = \dfrac{\gm_{\rm exc}}{\gm_{\rm exc;0}}
 \times
 }
 \nn\\
 && \dfrac{\Gm_{\rm rad}/\Gm_{{\rm rad};0}}{\Gm_{\rm rad}/\Gm_{{\rm rad};0} + \Gm_{\rm nrad}/\Gm_{{\rm rad};0} + (1-q_0)/q_0 } \dfrac{1}{q_0},
 \label{eq:bff}
\end{eqnarray}
requires a delicate balance of $\gamma_{\rm exc}$ and the radiative, $\Gm_{\rm rad}$, and nonradiative, $\Gm_{\rm nrad}$, decay rates~\cite{Anger2006,Bharadwaj2007,Ringler2008,Arruda2017a,Sun2017b,Sun20JPCC}.
Here the subscript ``0'' indicates the respective quantity in the free space, and $q_0$ is the intrinsic quantum yield.
$\Gm_{\rm rad}$ is determined by the local density of states, whereas $\Gm_{\rm nrad}$ is determined by metal losses. 
Assuming randomly oriented emitters, the decay rates in \eqref{eq:bff} are averaged over dipole orientation as $\Gm_{\rm nrad;rad} = ( \Gm^{\perp}_{\rm nrad;rad} + 2\Gm^{\parallel}_{\rm nrad;rad})/3$, where the superscripts ``$\parallel$'' and ``$\perp$'' denote the perpendicular (radial) and parallel (tangential) dipole orientation relative to the surface of plasmonic nanostructure.
The excitation rate in \eqref{eq:bff} is \textit{averaged} over the particle surface: $\gamma_{\rm exc} \propto \langle|\vE|^2\rangle$~\cite{Rasskazov19JOSAA,Sun20JPCC}. 
In general the Stokes shift is allowed between excitation, $\ld_{\rm exc}$, and emission, $\ld_{\rm ems}$, wavelengths, corresponding to the respective excitation ($\gamma_{\rm exc}/\gamma_{{\rm exc};0}$) and emission ($q/q_0$) processes.

Metal@dielectric core-shell nanoparticles~\cite{Moroz2005,Moroz2005a,Tovmachenko2006,Aslan2007,Reineck2013,Lu2014a,Wang2014a,Pang2015,Planas2016,Walters2018,Niu2018,Camacho2016,Meng2018,Wan2021,Sun20JPCC,Fontaine2020} possess unique advantages owing to their mass production capability with low cost chemical synthesis methods. 
Due to inextricably intertwined quantities of excitation, radiative, and nonradiative decay rates in \eqref{eq:bff}, previous studies were largely guided by the paradigm of {\em metal-enhanced fluorescence} (MEF)~\cite{Lakowicz2008}, which states that there is some optimal distance from a metallic surface where $\bar F$ attains its maximum. This arises because, by approaching a metal surface, $\Gm_{\rm rad}$ begins to increase sooner but $\Gm_{\rm nrad}$ increases faster. As a rule, it has been considered largely unfavourable to decrease plasmon coupling between the metal and fluorophore by increasing emitter separation from a metal surface by more than a couple of tens of nanometers. 
Therefore, previous studies~\cite{Tovmachenko2006,Sun20JPCC,Wan2021} have limited shell thickness, $t_s$, to $t_s\le 30$~nm, yielding for Au@dielectric core-shells a maximum achievable averaged $\bar F\approx 9$ ($\approx 70$ for Ag core) for shell refractive index $n_s\le 2$ \cite{Sun20JPCC}.

Importantly, identical core-shell under identical conditions may yield wildly differing values of $\bar F$ depending on the value of emitters $q_0$.
The second fraction on the rhs of \eqref{eq:bff} remains essentially constant whenever $q_0$ is varied over an interval such that the contribution $(1-q_0)/q_0$ is negligible compared to at least one of the other two terms in the denominator (e.g., $\Gamma_{\rm nrad}/\Gamma_{{\rm rad};0}$ or $\Gamma_{\rm rad}/\Gamma_{{\rm rad};0}$), which is rather common situation in the MEF~\cite{Moroz2005a,Anger2006,Sun20JPCC}.
Consequently, due to the third fraction on the rhs of \eqref{eq:bff}, the resulting fluorescence enhancement $\bar F\propto 1/q_0$: the smaller $q_0$ the larger $\bar F$.
Therefore, $\bar F$ is {\em ambiguous} when characterizing the environment of an emitter, unless, for the sake of comparison,
one agrees to fix the value of $q_0$. It is clear that the value of $q_0=1$ is the least favourable for $\bar F$.
Moreover, we rely on \textit{orientationally averaged} electric field intensity, $\langle |\vE|^2 \rangle$, being well aware that larger intensity values are available in hot spots~\cite{Wan2021}, in which $\bar F$ of an emitter could be easily increased by an order of magnitude compared to our averaged values.
Nonetheless, it is the value of $q_0=1$ and averaged (over the spherical surface) $\gm_{\rm exc} \propto \langle |\vE|^2 \rangle$ to which we stick throughout this work, very much as in our earlier work~\cite{Sun20JPCC}. 
Therefore, all the values of $\bar F$ reported below can be rather seen as the {\em lower} bound, or the \textit{worst case scenario} on the achievable enhancement, because the latter can be easily increased by decreasing $q_0$ or, by selective placing of emitter in a hot spot.

In this Letter, we show that by extending the parameter range for the core radius, $r_c$, and, importantly, the shell thickness, $t_s$, well beyond that suggested by MEF reasoning, one can obtain $\bar F$ of at least two orders of magnitude larger than for the MEF range, all that for experimentally feasible designs using common fluorophores.
An essential prerequisite for our simulations is recently reported efficient determination of orientationally averaged electric field intensities~\cite{Rasskazov19JOSAA,Rasskazov20OSAC}, supplementing earlier efficient calculation of decay rates~\cite{Moroz2005}~\footnote{its Fortran version for decay rates has been since 2006 freely available from 
\href{www.wave-scattering.com/codes.html}{www.wave-scattering.com/codes.html}}.
In essence, surface integrals of $|\vE|^2$ can be performed analytically~\cite{Rasskazov19JOSAA} and the calculation of average intensity costs the same computational time as determining intensity at a given point. The full power of recent analytic~\cite{Moroz2005,Rasskazov19JOSAA} and numerical~\cite{Rasskazov20OSAC} advances has allowed for unprecedented large-scale optimization studies~\cite{Sun20JPCC} that have been so far outside the reach of numerical methods. Unlike previous works on MEF~\cite{Sun20JPCC,Wan2021}, we have here significantly extended the parameter space for the metal core radius, $r_c$, and dielectric shell, $t_s$: $r_c \in [20,180]$~nm and $t_s \in [10,300]$~nm, keeping the overall size of core-shell particle less than a micrometer, much smaller than of typical large microresonators required to harness whispering gallery modes (WGMs)~\cite{Campillo1991,Barnes1996,Reynolds2017}.

%%%%%%%%%%%%%%%%%%%%%%%%%%%%%%%%%%%%%%%%%%%%%%%%%%%%%%%%%%%%%%%%%%%%%
\begin{figure}[t]
 \centering
 \includegraphics{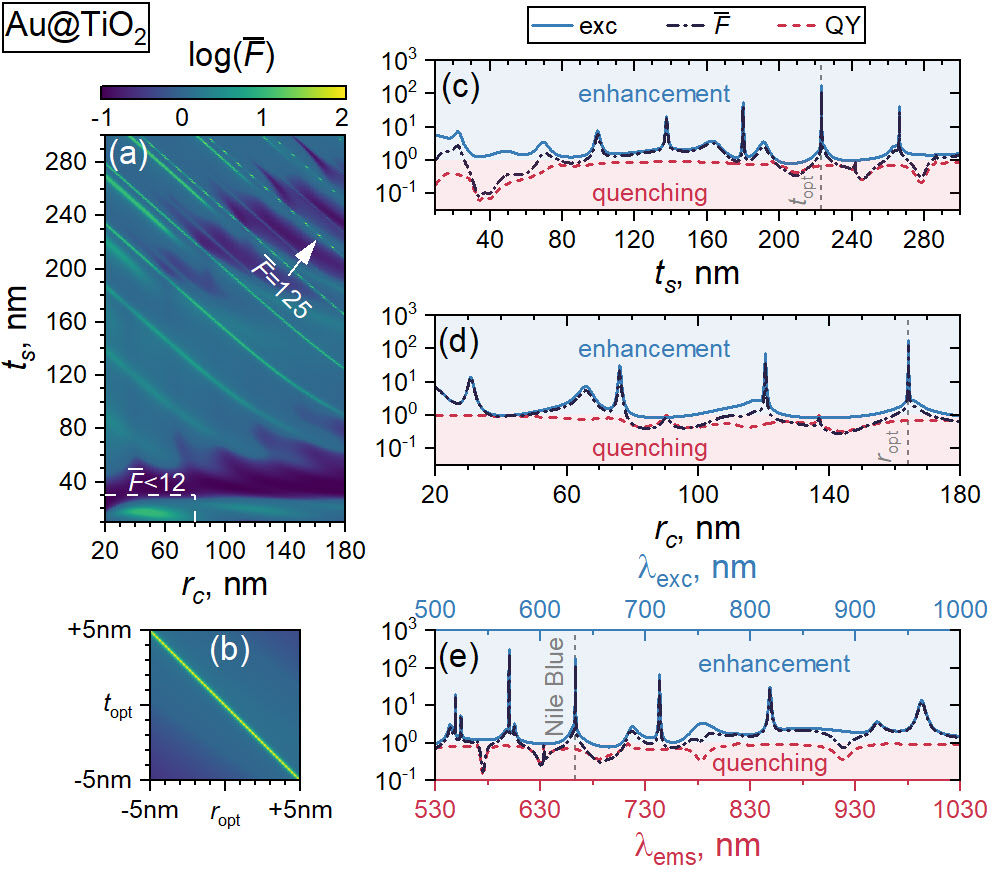}
 \caption{(a) Fluorescence enhancement $\bar F$ for Au@TiO$_2$ core-shell particles ($n_s=2.7$) embedded in air host ($n_h = 1$) as a function of shell thickness $t_s$ and core radius $r_c$ for Nile Blue dye with $\ld_{\rm exc} = 633$~nm and $\ld_{\rm ems}=663$~nm.
 The dashed rectangle in (a) shows the typical sizes of core-shells considered in the literature for the optimization problem~\cite{Sun20JPCC,Wan2021}.
 (b) $\bar F$ near the point highlighted by arrow in (a) within $\pm 5$~nm deviation from the optimal values $r_{\rm opt}=164.1$~nm and $t_{\rm opt}=223.1$~nm.
 Excitation (exc), quantum yield (QY) and fluorescence enhancements ($\bar F$) as functions of: 
 (c) shell thickness with fixed $r_c=r_{\rm opt}$, 
 (d) core radius with fixed $t_s=t_{\rm opt}$ and 
 (e) wavelength for a dye with $\ld_{\rm ems}-\ld_{\rm exc}=30$~nm Stokes shift.}
 \label{fig:AuTiO2}
\end{figure}
%%%%%%%%%%%%%%%%%%%%%%%%%%%%%%%%%%%%%%%%%%%%%%%%%%%%%%%%%%%%%%%%%%%%%
%%%%%%%%%%%%%%%%%%%%%%%%%%%%%%%%%%%%%%%%%%%%%%%%%%%%%%%%%%%%%%%%%%%%%
\begin{figure}[t]
 \centering
 \includegraphics{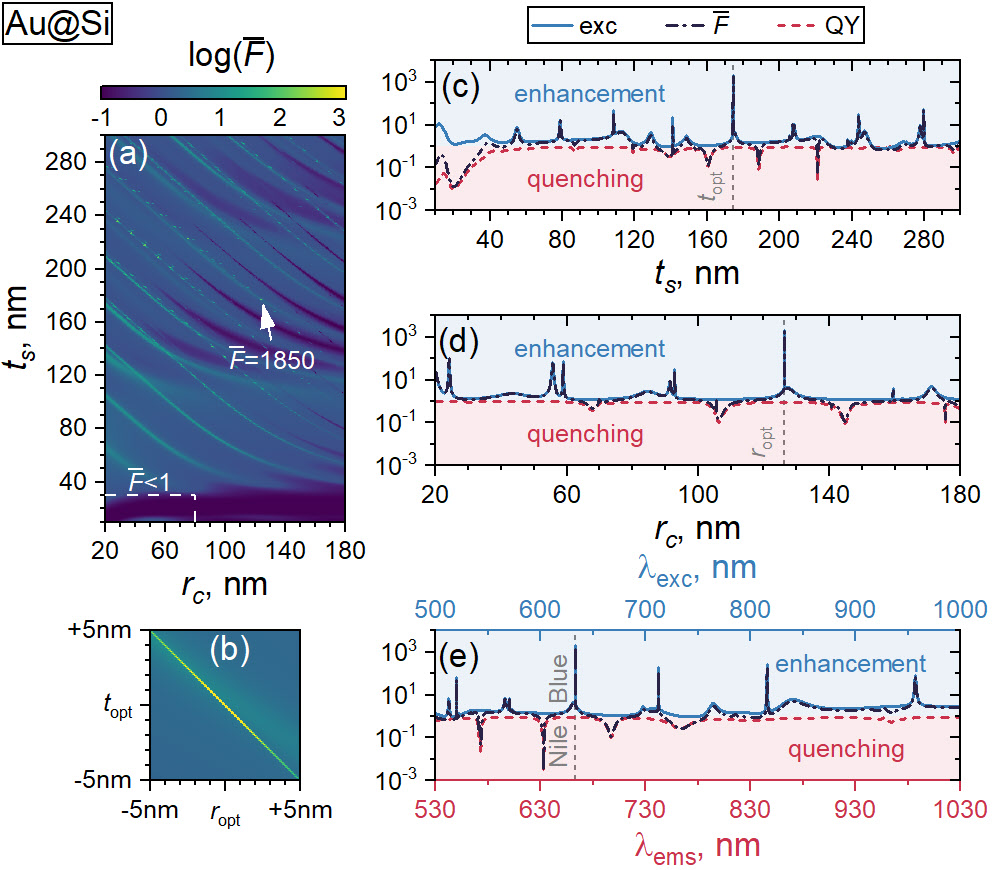}
 \caption{Analogous as in Fig.~\ref{fig:AuTiO2}, but for Au@Si core-shell particles ($n_s=3.5$) with $r_{\rm opt}=126.3$~nm and $t_{\rm opt}=174.2$~nm.}
 \label{fig:AuSi}
\end{figure}
\begin{figure}[t]
 \centering
 \includegraphics{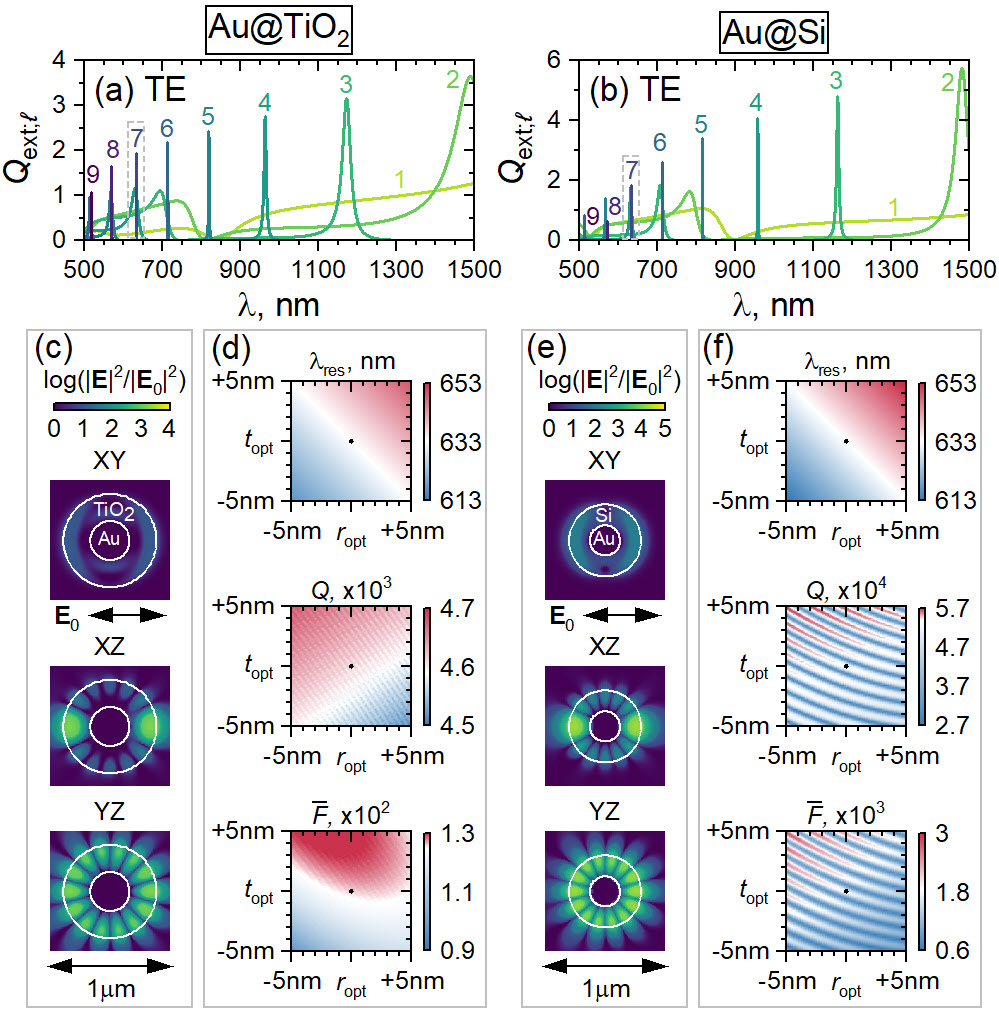}
 \caption{(a),(b) Multipole decomposition of the extinction efficiency $Q_{{\rm ext};\ell}$ for transverse electric, TE~\cite[Eq. (19)]{Rasskazov20OSAC}, polarization for Au@TiO$_2$ and Au@Si core-shells.
 Respective values of $\ell$ are shown in plots.
 (c),(e) Electric field intensities for a plane wave excitation along the $z$-axis and the field orientation along the $x$-axis at $\ld_{\rm exc}=633$~nm.
 (d),(f) wavelength, $\ld_{\rm res}$, and $Q$-factor of the $\ell=7$ TE resonance and 
 respective $\bar F$, similar as to Figs.~\ref{fig:AuTiO2}(b) and ~\ref{fig:AuSi}(b), but with $\ld_{\rm exc}=\ld_{\rm res}$.
 White color in colormaps corresponds to (d) $\bar F=125$, $Q=4.5\cdot 10^3$ and (f) $\bar F=1850$, $Q=5.4\cdot 10^4$}
 \label{fig:ext}
\end{figure}

%%%%%%%%%%%%%%%%%%%%%%%%%%%%%%%%%%%%%%%%%%%%%%%%%%%%%%%%%%%%%%%%%%%%%
We consider the core-shell nanoparticle with a gold core (with refractive index $n_c$~\cite{Johnson1972}) surrounded by a dielectric shell (with $n_s$ either $2.7$ for TiO$_2$ or $3.5$ for Si).
The nanoparticle is embedded in a homogeneous medium with a refractive index $n_h$, which is set to be air ($n_h=1$).
A fluorescence emitter (Nile Blue dye with $\ld_{\rm exc} = 633$~nm and $\ld_{\rm ems}=663$~nm) is modeled as an oscillating dipole located at $t_s+0.75$~nm distance from the metal surface.
In order to highlight the parameter range investigated in present work, the thin shell region examined in previous works~\cite{Tovmachenko2006,Arruda2017a,Sun20JPCC,Wan2021} essentially corresponds to the dashed rectangles 
shown in the bottom-left corner of Figs.~\ref{fig:AuTiO2}(a), ~\ref{fig:AuSi}(a). 
The metal@dielectric core-shell nanoparticles with sizes bounded with this dashed rectangles is what one would have investigated when guided by the MEF paradigm~\cite{Lakowicz2008}.
Small values of $\bar F\to 1$ within the highlighted dashed rectangle in Fig.~\ref{fig:AuSi}(a) are in line with results reported in Ref.~\cite{Wan2021} (Fig. 5 therein) that the maximum achievable $\bar F$ decreases when increasing $n_s$ above $2.6$.
As demonstrated in Figs.~\ref{fig:AuTiO2}(a) and \ref{fig:AuSi}(a) for Au@TiO$_2$ and Au@Si core-shells, respectively, essential for harnessing exceptionally large $\bar F$ is to have large field enhancement at the excitation wavelength, $\ld_{\rm exc}$, which can be seen in Figs.~\ref{fig:AuTiO2}(c)-(e) and \ref{fig:AuSi}(c)-(e).
A closer look reveals that the optimal locations correspond to higher-order $\ell=7$ transverse electric (TE) multipole resonance, see Fig.~\ref{fig:ext}(a)-(c),(e). 
The resonance is characterized by large values of its quality factor, $Q=\ld_{\rm res}/\Dt\ld_{\rm res}$ ($Q\sim 4.5\cdot 10^3$ for Au@TiO$_2$ and $Q\sim 5.4\cdot 10^4$ for Au@Si), where $\Dt\ld_{\rm res}$ is the corresponding full width at half maximum (FWHM) of the resonance. 
Along with thick shell providing a sufficient spacing from metal core for preventing quenching (i.e. $\Gm_{\rm nrad}/\Gm_{{\rm rad}}\ll 1$), the 
TE $\ell=7$-pole resonances in Figs.~\ref{fig:AuTiO2}(c) and \ref{fig:AuSi}(c) provide conditions for extraordinary large $\bar F=125$ and $\bar F=1850$ for Au@TiO$_2$ and Au@Si core-shells, respectively.
Our optimal configurations are obviously far away from the usual MEF range, because the $Q$-factors of the $\ell=7$-pole TE resonances are comparable to the highest $Q$-factors of purely dielectric structures of similar size~\cite{Huang2021}, both being between the $Q$-factors of typical plasmonic structures ($Q\lesssim 10^2$~\cite{Stockman2011}) and WGMs in large homogeneous silica spheres ($Q\ge 10^8$~\cite{Braginsky1989}). 

%%%%%%%%%%%%%%%%%%%%%%%%%%%%%%%%%%%%%%%%%%%%%%%%%%%%%%%%%%%%%%%%%%%%%
%%%%%%%%%%%%%%%%%%%%%%%%%%%%%%%%%%%%%%%%%%%%%%%%%%%%%%%%%%%%%%%%%%%%%
As expected, experimental imperfections in each of $r_c$ and $t_s$ within $\pm 5$~nm, influence the position and $Q$-factor of the relevant TE $\ell=7$-pole resonance (Fig.~\ref{fig:ext}(d),(f)). 
Importantly, within the above $\pm 5$~nm tolerances, the resonance position still remains within the dye excitation band, while $Q$-factor remains over $10^3$ and $10^4$ for Au@TiO$_2$ and Au@Si, respectively. 
As a rule, an emitter excitation band is typically much broader than the reported resonance FWHM, which in turn is much broader than laser diode linewidth that can be as narrow as 1 Hz \cite{Stoehr2006}.
Therefore, the above experimental imperfections can be easily accommodated with a tunable laser diode excitation source.
According to Figs.~\ref{fig:AuTiO2}(b) and \ref{fig:AuSi}(b), crucial for preserving high $\bar F$ in the case of fixed $\ld_{\rm exc}$ is to keep the total sphere radius, $r_c+t_s$, as close as possible to that of the optimal configuration.
When excitation wavelength tracks resonance wavelength, i.e. $\ld_{\rm exc}=\ld_{\rm res}$, Figs.~\ref{fig:ext}(d),(f) reveal that not only larger $\bar F\gtrsim 3000$ are in principle possible, but also parameter range for obtaining large values of $\bar F$ is broader than with fixed $\ld_{\rm exc}$.
%%%%%%%%%%%%%%%%%%%%%%%%%%%%%%%%%%%%%%%%%%%%%%%%%%%%%%%%%%%%%%%%%%%%%
\begin{figure}[t]
 \centering
 \includegraphics{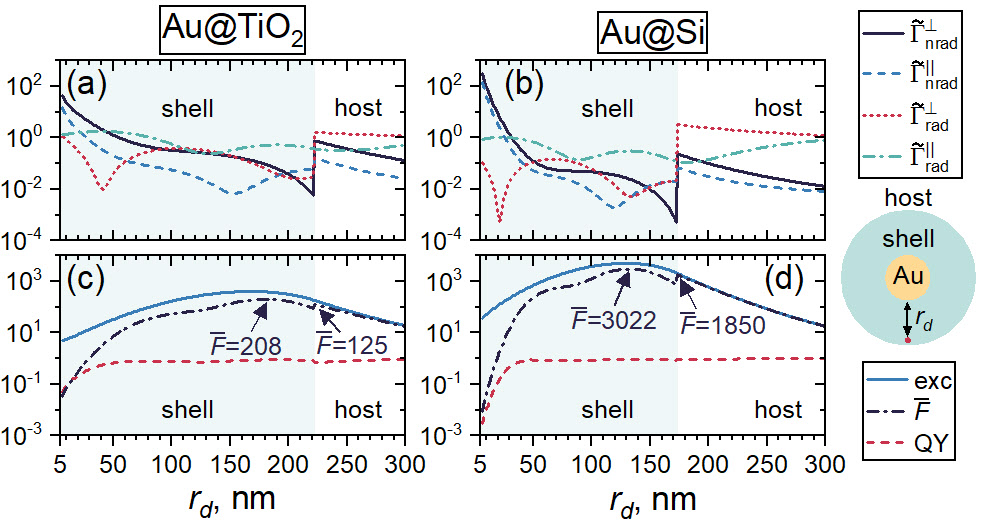}
 \caption{(a),(b) Radiative and nonradiative decay rates, $\tilde{\Gm}=\Gm/\Gm_{{\rm rad};0}$, normalized to the radiative decay rates in infinite homogeneous medium having the refractive index of the shell where the dipole emitter is located~\cite[Eqs.(28)-(29)]{Rasskazov20OSAC} and 
 (c),(d) excitation (exc), quantum yield (QY) and fluorescence enhancements ($\bar F$) as functions of the dipole position, $r_d$, within the shell and in the host (see sketch on the right) for the optimal configurations (see Figs~\ref{fig:AuTiO2} and \ref{fig:AuSi}) of Au@TiO$_2$ (left) and Au@Si (right) core-shells.}
 \label{fig:pos}
\end{figure}
%%%%%%%%%%%%%%%%%%%%%%%%%%%%%%%%%%%%%%%%%%%%%%%%%%%%%%%%%%%%%%%%%%%%%

Figure~\ref{fig:pos} makes it transparent that a nanometer precise experimental control of placing dipole emitters at a predetermined radial position of a dielectric shell~\cite{VanBlaaderen1992} may enable even larger values of $\bar F$ compared to those on the shell surface: $208$ and $3022$ for Au@TiO$_2$ and Au@Si core-shells, respectively. 
Moderate values of decay rates imply that the MEF scaling, $\bar F\propto 1/q_0$, no longer holds in our case. 
Interestingly, apart a near proximity of metal core, $\bar F$ at the optimal configurations nearly exactly follows the excitation enhancement as the dipole position is varied. 
Nearly continuous $\gamma_{\rm exc}$ across the shell-host interface is a consequence of that the field intensity is largely dominated by the tangential components of $\vE$. Contrary to that, for thin shells one observes a clear discontinuous drop of $\gamma_{\rm exc}$ in optically denser shell~\cite[Fig. S1]{Sun20JPCC} as a consequence of $n_s E_s^\perp=n_h E_h^\perp$ satisfied by the normal components $\vE$ at the interface. 

A natural question appears: {\em Given that the shell is thick, does one need at all a metallic core?} The answer is one really does, because metal constituents are in principle required for large $\bar F$~\cite{Bozhevolnyi2016}, and nothing comparable happens for analogous TE resonances in a comparable homogeneous dielectric spheres~\cite{DeDood2001a}. 
However, because of large shell thickness, and thus large dye-metal surface separation,
the term ``metal-enhanced fluorescence'' acquires completely different meaning, because metal plays a more subtle role in the extraordinary fluorescence reported here. 
The latter can be visualized as that metal core plays the role of a reflecting surface pushing the electric field intensity from the core interior into the shell, when compared to the case of a homogeneous dielectric sphere. 
In intuitive terms, an optimal shell appears to require an effective optical thickness, $L=[4(r_s^3 - r_c^3)] /[3(r_s^2 + r_c^2)]$, which corresponds to the mean-free path of excitation waves within the shell~\cite{Moroz2008}, to be $(\ell \ld_{\rm exc})/(4n_s)$, where $r_s = r_c + t_s$.
Combined with the phase shift $\pi$ upon reflection on the metal core surface, the excitation light arriving at the dipole position is then in phase with that being reflected at the metal core surface, leading to a constructive interference, and the extraordinary fluorescence.
Although the phase shift $\pi$ upon reflection occurs only for a perfect metal~\cite{Doi1997}, we find, nevertheless, the above intuitive argument, which can be seen as an adaption of the ``$\ld/4$'' condition in radiowave antennas to core-shell geometry, surprisingly precise: at the above reported extraordinary values of $\bar F$ one has $(4Ln_s)/(\ld_{\rm exc}\ell) \approx 0.98$ and $(4Ln_s)/(\ld_{\rm exc}\ell) \approx 0.99$ for Au@TiO$_2$ and Au@Si core-shells, respectively.
This argument explains why relatively high $n_s$ is important for keeping the core-shell particle diameter below $1 \mu$m and why optimal $t_s$ is thinner for Si shell than for TiO$_2$ shell.

%%% Discussion

The values of (non-averaged) $F\sim 10^3$ have been reported earlier~\cite{Kinkhabwala2009,Hoang2015}.
A closer inspection shows, however, that they were obtained in the MEF regime for an organic dye with low $q_0$ ($q_0 = 0.025$ in Ref.~\citenum{Kinkhabwala2009} and $q_0=0.1$ in Ref.~\citenum{Hoang2015}). 
An equivalent result for dyes with $q_0=1$, which should be compared with our result, yields however mere $F \sim 10^2$ after rescaling. 
The only exception is recent use of fluorescent dye Atto 532 with $q_0=0.9$ in Ref.~\citenum{Traverso2021} providing $F=910$.
Finally, for a complementary case of a dielectric@metal core-shell, the highest fluorescence enhancements were observed at the Fano resonance for a dielectric ($n_c=3.5$) core and metal (Ag) shell~\cite{Arruda2017a}: $F^\perp=50$ and $F^\parallel=2.5$ in the case of radially and tangentially oriented dipole located at a hot spot, yielding the orientation-averaged enhancement of mere $F\approx 18$.

To conclude, by scanning over more than $4\cdot 10^6$ configurations of core-shell Au@dielectric nanoparticles, we observed configurations with average fluorescence enhancement as high as $\gtrsim 3000$ on the shell surface or in its interior. 
Actual fluorescence enhancements can be increased further by taking advantage of hot spots.
Surprisingly, the extraordinary values of $\bar F$ were obtained for conventional metal@dielectric core-shell nanoparticles, without any need of a fancy nontrivial shape, such as bowtie nanoantenna~\cite{Kinkhabwala2009} or a nanocube~\cite{Hoang2015,Traverso2021}.
In order to investigate a realistic case, we have selected Nile Blue dye as a model case. 
It is obvious that any other suitable electric dipole point-like emitters, such as different dyes, quantum dots, diamond impurities etc. can be used, with the core-shell structure being adjusted to their excitation and emission wavelengths.

Our results suggest a paradigm shift from the conventional ``metal-enhanced fluorescence'', whereby the ``metal'' is now becoming rather auxiliary than determinative constituent. 
Therefore, analogous results are expected for other good reflecting metal (e.g. Ag, Al, Mg) cores, as long as the shell refractive index is sufficiently high. TiO$_2$ and Si have been used here merely as nonlimiting examples of shell dielectric materials with high refractive index ($2.7$ for TiO$_2$ and $3.5$ for Si). 
We anticipate that averaged $\bar F\gtrsim 100$ can be routinely obtained with $n_s\gtrsim 2.2$, which can be achieved with common Ta${}_2$O${}_5$, ZnS, and Nb${}_2$O${}_5$ coatings.
Our results have immediate applications in biological sensing~\cite{Zhang2017c}, lasing~\cite{Mamonov2018} (including spasers~\cite{Bergman2003}), cell tagging~\cite{Humar2017}, and are suggesting the upconversion~\cite{Fernandez-Bravo2018,Liu2020d} enhancements of $\gtrsim 10^4$. Exceptionally high field intensities within the shell beg for their exploitation in various nonlinear applications. We hope that our results could stimulate experimental works on synthesis of metal@dielectric core-shells with high-index shells~\cite{Mohapatra2008} or analogous investigation of other particle shapes~\cite{Abadeer2014,Pavelka2021}.

\textbf{Disclosures.} The authors declare no conflicts of interest.

%%%%%%%%%%%%%%%%%%%%%%%%%%%%%%%%%%%%%%%%%%%%%%%%%%%%%%%%%%%%%%%%%%%%%

%%%REFERENCES%%%
% \bibliography{references}
% \bibliographyfullrefs{references}

\end{document}